\documentclass[twocolumn, showpacs, prl]{revtex4}

\usepackage{graphicx}% Include figure files

\usepackage{dcolumn}% Align table columns on decimal point

\usepackage{bm}% bold math

\begin{document}

\title{Spin-dependent resistivity at transitions between integer quantum Hall states}

\date{\today}

\author{K. Vakili}

\author{Y. P. Shkolnikov}

\author{E. Tutuc}

\author{N. C. Bishop}

\author{E. P. De Poortere}

\author{M. Shayegan}

\affiliation{Department of Electrical Engineering, Princeton
University, Princeton, NJ 08544}

\begin{abstract}

The longitudinal resistivity at transitions between integer
quantum Hall states in two-dimensional electrons confined to AlAs
quantum wells is found to depend on the spin orientation of the
partially-filled Landau level in which the Fermi energy resides.
The resistivity can be enhanced by an order of magnitude as the
spin orientation of this energy level is aligned with the majority
spin. We discuss possible causes and suggest a new explanation for
spike-like features observed at the edges of quantum Hall minima
near Landau level crossings.

\end{abstract}


\maketitle

The integer quantum Hall (QH) effect is a relatively
well-understood phenomenon in condensed matter physics that occurs
when a two-dimensional electron system (2DES) is subjected to a
large perpendicular magnetic field \cite{klitzing80}. Associated
with the effect are oscillations in the longitudinal
magnetoresistivity (MR), $\rho$$_{xx}$, including wide regions of
zero resistivity, that occur as the Fermi level passes through
successive disorder-broadened Landau levels (LLs) containing bands
of localized and extended states \cite{prange90}. At transitions
between integer QH states, there are peaks in $\rho$$_{xx}$ that
have been extensively studied in the context of universal scaling
phenomena \cite{huckestein95}.  The magnetic field resolves the
LLs into up and down spin branches split by the Zeeman energy,
causing the $\rho$$_{xx}$ maxima to split as well, but the spin
degree of freedom is generally thought to play a minor role in
determining the value of the $\rho$$_{xx}$ maxima in the
spin-resolved QH regime.

We observe unexpectedly large changes of $\rho$$_{xx}$ at
transitions between integer QH states as the 2DES is tilted with
respect to an applied magnetic field such that the LL spins are
polarized (Fig. 1). Specifically, the value of $\rho$$_{xx}$ at a
transition is higher when the spin of the energy level in which
the Fermi energy ({\it E}$_{F}$) resides is aligned with the
majority, lowest LL spin, and lower for those levels with the
opposite spin orientation. The resistivity difference can be as
large as an order of magnitude.

We performed measurements on electrons confined to a narrow
(45$\AA$ wide) AlAs quantum well (QW) with Al$_{0.4}$Ga$_{0.6}$As
barriers, grown on a GaAs (100) substrate. The 2DES in this system
occupies a single, out-of-plane valley with an isotropic Fermi
contour, a band effective mass of 0.21{\it m}$_{e}$, and a band
g-factor of 2. We also made measurements on 2D electrons in a
strained, wide (110$\AA$ wide) AlAs QW, where a single, in-plane
valley with an anisotropic Fermi contour with transverse and
longitudinal band effective masses of 1.1{\it m}$_{e}$ and
0.21{\it m}$_{e}$ is occupied. We studied a total of four samples
from three different wafers, all of which were lithographically
patterned in a Hall bar configuration with Ohmic contacts made by
depositing AuGeNi and alloying in a reducing environment. Metallic
front and back gates were deposited for {\it in situ} control of
the charge density, {\it n}. The peak mobilities are 5.0
m$^{2}$/Vs in the narrow QW samples and 17 m$^{2}$/Vs for the
low-mass direction of the wide QW. We performed measurements down
to {\it T} = 20 mK using low-frequency lock-in techniques. The
samples were mounted on a single-axis tilting stage to allow the
angle, $\theta$, between the normal to the sample and the magnetic
field to be varied from 0$^{o}$ to 90$^{o}$ {\it in situ}.  We
define {\it B}$_{tot}$ as the total magnetic field, and {\it
B}$_{\|}$ and {\it B}$_{\bot}$ as the components parallel or
perpendicular to the 2DES plane.

Because the Zeeman energy, {\it E}$_{Z}$, depends on {\it
B}$_{tot}$ while the cyclotron energy, {\it E}$_{c}$, depends on
{\it B}$_{\bot}$, the ratio {\it E}$_{Z}$/{\it E}$_{c}$ can be
adjusted by changing $\theta$ \cite{fang68}. At the so-called
coincidence angles, {\it E}$_{Z}$/{\it E}$_{c}$ takes integer
values as different spin branches of the various LLs are brought
into energetic coincidence.  These level crossings can be labelled
by a crossing index given by the difference between the LL indices
for the crossing up and down spin levels, {\it i} = {\it
N}$_1^{\downarrow}$ - {\it N}$_2^{\uparrow}$. By tilting the
sample through coincidence angles, the spin for a LL associated
with a given $\rho$$_{xx}$ maximum can be reversed [see Fig.
2(a)]. This argument is based on a single-particle picture and, as
discussed later, many-body effects can alter this simple behavior
near the coincidence angles.

AlAs QWs are very well suited to tilted-field measurements. Their
relatively large value of {\it g}*{\it m}* \cite{vakili04,
shkolnikov04} ({\it g}* and {\it m}* are the renormalized
Land\'{e} g-factor and electron mass respectively) allows LL
coincidences to be reached at modest tilt angles as compared, for
example, to GaAs 2DESs which have an order of magnitude smaller
{\it g}*{\it m}*. In narrow AlAs QWs, {\it g}*{\it m}* is small
enough to allow the {\it i} = 1 coincidence to be observed, while
wide AlAs QWs are past this coincidence at $\theta$ = 0$^{o}$ for
most accessible densities \cite{shkolnikov04, depoortere00}.
Because of their relatively narrow well widths, AlAs QWs have the
further advantage of being largely free of finite-thickness
effects associated with the coupling of {\it B}$_{\|}$ to the
orbital degree of freedom.

\begin{figure}
    \centering
    \includegraphics[scale=0.6]{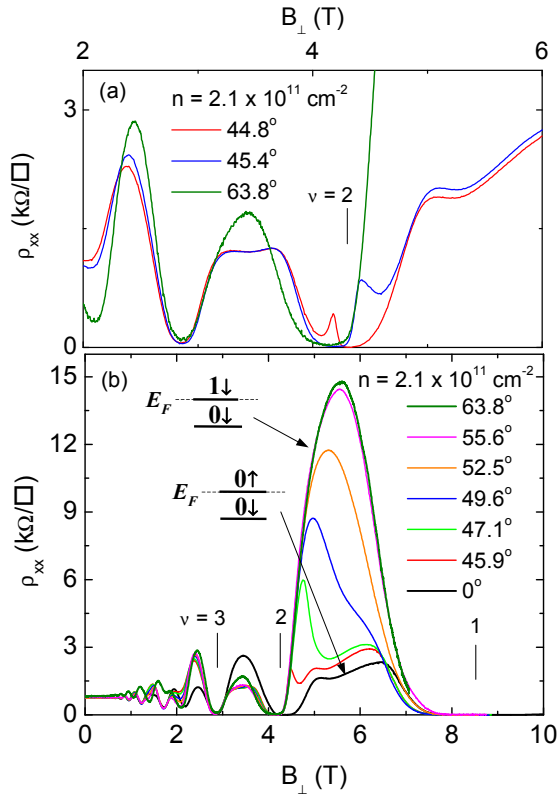}
    \caption{(color online) $\rho$$_{xx}$ vs. {\it B}$_{\bot}$ at {\it T} = 300 mK in a narrow AlAs QW as the sample is tilted through
    the first ({\it i} = 1) coincidence angle ($\theta$ $\simeq$ 45$^o$).  (a) A resistance spike is visible
    near the $\nu$ = 2 minimum, and (b) subsequently merges with the $\rho$$_{xx}$ maximum
    between $\nu$ = 2 and 1, leading to a dramatic enhancement of $\rho$$_{xx}$ at that maximum.  The two
    lowest energy levels, labelled by their LL index and spin, and the location of {\it E}$_{F}$
    are shown schematically for the indicated $\rho$$_{xx}$ maxima.}
\end{figure}

In Figs. 1(a) and (b), we show $\rho_{xx}$ in a narrow QW at
several different $\theta$. The value of {\it g}*{\it m}* is about
1.5 at {\it n} = 2.1 x 10$^{11}$ cm$^{-2}$, and previous
measurements have indicated no strong dependence of {\it g}*{\it
m}* on magnetic field or filling factor ($\nu$) in this system
\cite{vakili04}. This means that {\it E}$_{Z}$ $<$ {\it E}$_{c}$
at $\theta$ = 0$^{o}$, so that the $\rho$$_{xx}$ maximum near {\it
B}$_{\perp}$ = 6 T corresponds to {\it E}$_{F}$ passing through
the 0$\uparrow$ energy level (transition between $\nu$ = 2 and 1
QH states) [see Fig. 1(b)]. As the sample is tilted to angles near
the {\it i} = 1 coincidence, a spike develops in the $\nu$ = 2
minimum [Fig. 1(a)]. Similar spikes were observed previously in
wide AlAs QWs \cite{depoortere00} and were associated with
scattering at boundaries between different spin domains
concomitant with a first-order, many-body spin reversal
\cite{jungwirth01}. The properties of the spikes observed in
narrow QWs are similar, including hysteretic behavior and the
movement of the spike to higher {\it B}$_{\bot}$ as $\theta$ is
increased. In contrast to previous studies, however, when the
spike eventually merges with the $\rho$$_{xx}$ maximum between
$\nu$ = 2 and 1 [Fig. 1(b)], the spike's height and width increase
continuously until it is no longer distinct from the maximum
itself, and we are left with a single, {\it enhanced}
$\rho$$_{xx}$ maximum. Even after the spike is unresolvable, the
$\rho$$_{xx}$ maximum continues to increase in height as the
sample is tilted until a saturation angle is reached. The apparent
position of the QH minimum also depends on the spin orientation of
the adjacent LLs. It is evident in Fig. 1 that $\rho$$_{xx}$ at
$\nu$ $\gtrsim$ 2 attains finite values at lower {\it B}$_{\bot}$
when the second lowest LL is 1$\downarrow$ than when it is
0$\uparrow$. This causes an apparent shift in the position of the
$\nu$ = 2 minimum to lower {\it B}$_{\bot}$ \cite{angle}. The Hall
resistance plateaus (not shown) are truncated in step with the
$\rho$$_{xx}$ minima.

\begin{figure}
    \centering
    \includegraphics[scale=0.49]{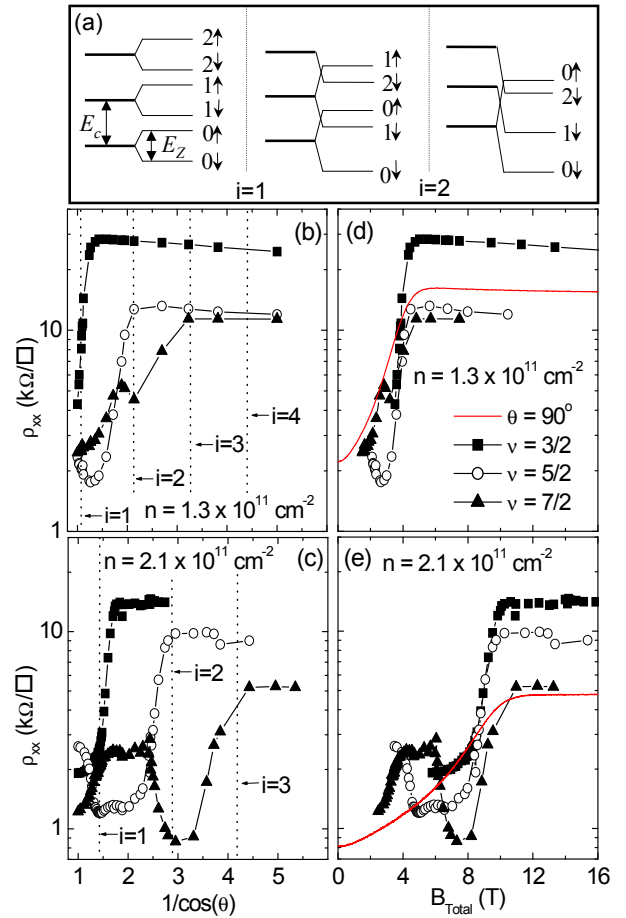}
    \caption{(color online) (a) Energy level configurations for a 2DES in a narrow AlAs QW as it is tilted in field.  Many-body effects
    near the coincidence angles are not shown. (b, c) $\rho$$_{xx}$ vs. 1/cos($\theta$) at {\it T} = 20 mK for
    half-integer $\nu$ at two densities, with the coincidence angles marked by vertical dotted lines.
    (d, e) The same data plotted against {\it B}$_{tot}$ along with parallel-field MR (red traces).  The key for the
    symbols in panels (b-e) is in panel (d).}
\end{figure}

To further understand the enhancement, we plot $\rho$$_{xx}$ at
half-integer $\nu$ in Fig. 2 as a function of $1/cos(\theta)$ [(b)
and (c)] and {\it B}$_{tot}$ [(d) and (e)] for two different {\it
n} in a narrow QW. We also show schematically the evolution of the
energy levels with increasing $\theta$ for the narrow QW in Fig.
2(a), although we have omitted peculiarities of the level
configurations associated with many-body effects near the
coincidence angles. Prior to the {\it i} = 1 coincidence,
$\rho_{xx}$ at $\nu$ = 3/2 and 7/2 is locally minimal as a
function of $1/cos(\theta)$ while at $\nu$ = 5/2 it is maximal.
From the energy level diagram in Fig. 2(a), we see that 3/2 and
7/2 correspond to {\it E}$_{F}$ passing through spin-up branches
while 5/2 corresponds to a down branch.  As the {\it i} = 1
coincidence is traversed, the spin for the energy levels
corresponding to these $\nu$ flip, and accordingly $\rho$$_{xx}$
increases or decreases as the spin flips to down or up
respectively.  With further increase of $\theta$, this
correspondence between $\rho$$_{xx}$ and spin continues, and
eventually $\rho$$_{xx}$ at a given $\nu$ saturates at the angle
beyond which there are no further spin reversals. Beyond this
saturation, there sometimes appears a slow decrease in
$\rho$$_{xx}$ (e.g. $\nu$ = 3/2 at {\it n} = 1.3 x 10$^{11}$
cm$^{-2}$), however this is small compared to the initial increase
and does not appear to depend on the spins of the nearby (higher)
LLs (i.e., it is unaffected by passage through subsequent
coincidence angles).

Plotting $\rho$$_{xx}$ at the various half-integer $\nu$ as a
function of {\it B}$_{tot}$ [Figs. 2(d,e)] reveals that the
enhancement saturates at a roughly $\nu$-independent {\it
B}$_{tot}$ which is close to the saturation field, {\it B}$_{P}$,
for the parallel-field MR where $\theta$ = 90$^{o}$ and {\it
B}$_{tot}$ = {\it B}$_{\|}$. {\it B}$_{P}$ has been shown
experimentally \cite{okamoto99, tutuc01} and theoretically
\cite{dolgopolov00} to correspond to the total spin polarization
of the charges. Thus, the correspondence between the saturation
fields, which is consistent with previous observations of isotropy
and field-independence of {\it g}*{\it m}* in narrow AlAs QWs
\cite{vakili04}, corroborates the assertion that the enhancement
of the $\rho$$_{xx}$ maxima are associated with spin polarization.
The analogy with parallel-field MR implies a possible explanation
for the occurrence of spin-dependent $\rho$$_{xx}$ in the QH
regime.

In the absence of orbital effects, the parallel-field MR can be
attributed largely to the spin-polarization dependence of
screening in a 2DES \cite{dolgopolov00}.  This is essentially a
consequence of the Pauli exclusion principle, whereby like-spin
charges screen disorder from each other less effectively than
opposite spin charges. This might seem irrelevant to the QH case
where the density of states is quantized into discrete (albeit
disorder-broadened), spin-resolved, and macroscopically degenerate
LLs. However, screening associated with inter-LL excitations can
indeed play a significant role in this regime \cite{smith92}. In
particular, considering that the relatively large band mass in
AlAs causes the electron-electron interaction energy, {\it
E}$_{e}$ $\sim$ {\it e}$^{2}$/4$\pi$$\epsilon${\it l}$_{B}$ ({\it
l}$_{B}$ is the magnetic length), to be greater than  {\it
E}$_{c}$ for all accessible fields, it becomes evident that mixing
between different LLs must be taken into account. Disorder may
then be screened most effectively when the spin of the LL in which
{\it E}$_{F}$ resides is different than the majority spin of the
low-lying LLs.

Treating different spin levels as parallel conduction channels
could also point to a cause of spin-dependent $\rho$$_{xx}$. The
spin-channel Hall resistivities ($\rho$$_{xy}^s$, s $\in$
$\lbrace$$\uparrow,\downarrow$$\rbrace$) are different by virtue
of the different filling factors for the two spin channels.  Thus,
if the spin-channel longitudinal resistivities ($\rho$$_{xx}^s$)
are close, then the spin-channel longitudinal conductivities,
$\sigma$$_{xx}^s$ = $\rho$$_{xx}^s$/($\rho$$_{xx}^s$$^2$ +
$\rho$$_{xy}^s$$^2$), are different and, when added and inverted,
give a spin dependent total $\rho$$_{xx}$. Of course,
$\rho$$_{xx}^\uparrow$ and $\rho$$_{xx}^\downarrow$ are not
necessarily close, and if the spin-channel conductivities are
calculated directly instead of the resistivities, as is usually
the case, then some additional mechanism must be postulated to
account for the spin-dependence. For $\rho$$_{xx}^\uparrow$ =
$\rho$$_{xx}^\downarrow$ $\ll$
$\rho$$_{xy}^{\downarrow,\uparrow}$, this model gives a factor of
9 change in $\rho$$_{xx}$ at $\nu$ = 3/2 when tilting past the
{\it i} = 1 coincidence, which is in reasonable agreement with the
narrow QW results. As we show below, however, measurements in wide
AlAs QWs give results that are not accounted for by this simple
model for spin-dependent resistivity.

\begin{figure}
    \centering
    \includegraphics[scale=0.43]{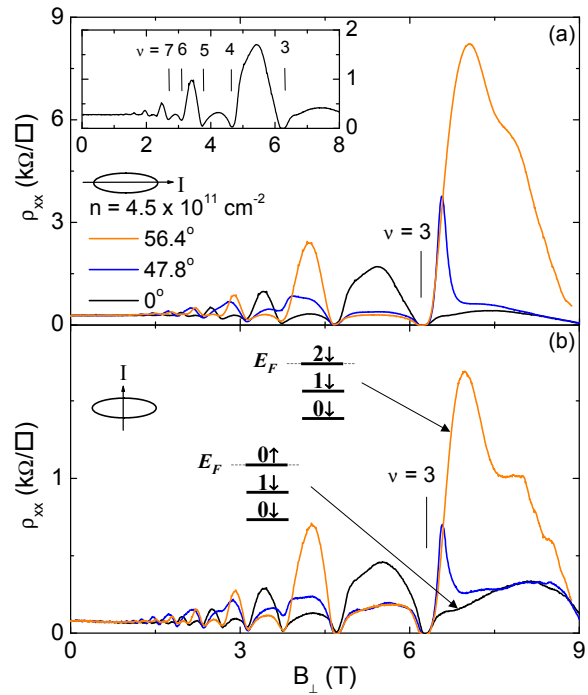}
    \caption{(color online) $\rho$$_{xx}$ vs. {\it B}$_{\bot}$ at {\it T} = 300 mK in a wide AlAs QW for different
    $\theta$ with the current ({\it I}) directed (a) along and (b) transverse to the long-axis of
    the occupied valley, as indicated schematically in each panel. The inset of (a) shows a closeup of
    the $\theta$ = 0$^o$ trace, and the LL indices and spins for the lowest three energy levels are shown in (b)
    for the indicated traces.}
\end{figure}

Our measurements of strained, wide AlAs QWs with a single
anisotropic valley occupied \cite{shkolnikov04b} give results that
are similar to those in the narrow QWs but with some notable
differences. Figure 3 displays MR data for current ({\it I})
oriented (a) along and (b) transverse to the long axis of the
occupied valley. At {\it n} = 4.5 x 10$^{11}$ cm$^{-2}$, the value
of {\it g}*{\it m}* in this system is about 3 \cite{shkolnikov04},
so the two lowest energy levels are spin-down and the highest
fully occupied levels are spin-up for odd integer $\nu$ $>$ 1 at
$\theta$ = 0$^o$. Like narrow QWs, the $\rho$$_{xx}$ maxima in the
wide QW are larger for {\it E}$_{F}$ passing through the majority
spin levels than through the minority spin levels. This is clearly
evident in wide AlAs QWs even at $\theta$ = 0$^o$ [Fig. 3(a)
inset], where the $\rho$$_{xx}$ maxima to the left of odd $\nu$
minima (spin-down levels) are large while those to the right
(spin-up levels) are small. In tilted field, resistivity spikes
form near the coincidence angles (the {\it i} = 2 coincidence at
{\it n} = 4.5 x 10$^{11}$ cm$^{-2}$ in this system occurs at
$\theta$ $\simeq$ 46$^o$) and large $\rho$$_{xx}$ changes occur
when these angles are passed [Figs. 3(a,b)]. In contrast to the
narrow QWs, however, the enhancement at $\nu$ = 5/2 in the wide QW
is anisotropic, with a remarkable factor of $\sim$19 enhancement
in (a) but only $\sim$5 in (b). By straining the sample to
depopulate one valley and populate the other and repeating the
measurement, we have confirmed that the enhancement anisotropy
arises from the valley (mass) anisotropy and not from the
orientation of {\it B}$_{||}$ with respect to {\it I} or the
valley. The narrow wells, which have an isotropic in-plane Fermi
contour, show no enhancement anisotropy. Such an anisotropic
enhancement is not explained by the aforementioned two-channel
model for spin-dependent $\rho$$_{xx}$. We also note that the
strong spin-dependence is suppressed when the sample is strained
to occupy a second valley. This might be expected in the context
of an inter-LL screening picture, since the additional valley
degree of freedom would allow like-spin charges to screen disorder
from each other as effectively as opposite-spin. That
spin-dependent $\rho$$_{xx}$ is strongest in single-valley systems
with large electron effective mass may explain why this effect has
not been reported for Si-MOSFETs (metal-oxide-semiconductor
field-effect transistors) or GaAs heterostructures \cite{emanuel}.

Whatever the cause, it is clear that $\rho$$_{xx}$ at QH
transitions is spin-dependent. This observation prompts a simple
explanation for $\rho$$_{xx}$ spikes that exist on the flanks of
QH states that is independent of domain-wall scattering. It has
been shown theoretically \cite{jungwirth01, giuliani85} that, when
opposite spin LLs are brought near energetic coincidence, {\it
B}$_{\bot}$ can induce a discontinuous transition where the nearly
coincident LLs suddenly trade places in energy. For example,at low
{\it B}$_{\bot}$ the energy levels may have the configuration of
the middle panel of Fig. 2(a), but as {\it B}$_{\bot}$ is
increased enough to begin depopulating the low lying LLs a rapid
transition to the configuration of the left panel of Fig. 2(a) can
occur. Spikes in $\rho$$_{xx}$ have been shown to accompany such
transitions \cite{depoortere00}, and have been explained as
arising from scattering at boundaries between domains of the two
competing energy level configurations \cite{jungwirth01}.  In the
case of a spin-dependent $\rho$$_{xx}$, however, such spikes would
accompany a collective spin-reversal transition at the edges of QH
states even in the absence of domain-wall scattering.  As an
example, for an appropriate $\theta$, the system may have the
configuration of the middle panel of Fig. 2(a) as {\it E}$_{F}$
enters the extended states of the 1$\downarrow$ energy level, and
because the spin of that LL is down, $\rho$$_{xx}$ would be large.
As {\it B}$_{\bot}$ is increased and a transition to the
configuration in the left panel of Fig. 2(a) is made, the
partially occupied LL would change to 0$\uparrow$ and
$\rho$$_{xx}$ would rapidly drop to the un-enhanced values. This
would register in $\rho$$_{xx}$ as a spike-like feature.

The spikes occurring at the edges of QH states are observed to
follow the same rising curve defined by the enhanced (spin-down)
$\rho$$_{xx}$ maxima (Figs. 1 and 3), which is consistent with
them originating from spin-dependent $\rho$$_{xx}$. Therefore, the
relative sizes of the spikes as they pass across the landscape of
$\rho$$_{xx}$ are neatly accounted for, as is the fact that the
largest spikes are observed when the spin-dependent enhancement is
strongest [e.g. Fig. 3(a)]. It is important to point out, however,
that resistivity spikes are also observed {\it deep within} the QH
minima (e.g. $\theta$ = 44.8$^o$ trace in Fig. 1(a)), i.e. in the
localized states, that do not fall within the envelopes defined by
the enhanced $\rho$$_{xx}$ maxima. Domain-wall scattering may
account for these, but we reiterate that large spike-like features
at the {\it edges} of QH states are an inevitable consequence of a
spin-dependent $\rho$$_{xx}$ with a sudden spin-reversal
transition. Spikes will occur even for second-order (continuous)
transitions as long as the transition is sufficiently rapid in
magnetic field on the scale of the QH features. In fact, the
sharpness of the transitions appears to diminish as they move
deeper into the extended states (i.e. deeper into the flanks of
the QH minima) [Fig. 1(b)], possibly signaling a crossover from a
first to a second-order transition. Finally, for spikes arising
from spin-dependent $\rho$$_{xx}$, the spike maximum may not
correspond to the center of the phase transition as was previously
thought to be the case, nor does the full-width-at-half-maximum
give the width of the transition. These facts place restrictions
on deriving information about the exchange energy or Curie
temperature associated with the ferromagnetic transition for
spikes occurring at the edges of QH states.

We thank the NSF for support, and A.\ H.\ MacDonald, R.\ N.\
Bhatt, S.\ L.\ Sondhi, F.\ D.\ M.\ Haldane, and R.\ Winkler for
discussions. Part of our work was performed at the Florida NHMFL,
also supported by the NSF; we thank E. Palm, T. Murphy, and G.
Jones for assistance.

\break


\begin{references}
\bibitem{angle}
We determined $\theta$ from Hall resistance and by matching
low-field, tilted-field $\rho$$_{xx}$ minima (those not
experiencing a LL crossing) to the same minima at $\theta$ =
0$^{o}$.
\bibitem{emanuel}
A spin-dependent $\rho$$_{xx}$ is evident, though not highlighted,
in \cite{tutuc01}.  Due to the large 2DES width, there is likely
also an orbital component to the tilted-field MR enhancement in
that measurement.
\end{references}
\end{document}